\documentclass[12pt,preprint,showpacs,superscriptaddress,pre]{revtex4-1}
\usepackage{amsmath}
\usepackage{amssymb}
\usepackage[dvips]{graphicx}
\usepackage[svgnames]{xcolor}
\usepackage[colorlinks=true,citecolor=blue,linkcolor=blue,citebordercolor=white,linkbordercolor=white]{hyperref}
\usepackage{natbib}

\def\eq#1{{Eq. (\ref{#1})}}

\begin{document}

\title{Diffusion of a particle in the spatially correlated exponential random energy landscape: transition from normal to anomalous diffusion}

\author{S.V. Novikov}
\email{novikov@elchem.ac.ru}
\affiliation{A.N. Frumkin Institute of
Physical Chemistry and Electrochemistry, Leninsky prosp. 31,
119071 Moscow, Russia}
\affiliation{National Research University Higher School of Economics, Myasnitskaya Ulitsa 20, Moscow 101000, Russia}



\begin{abstract}
Diffusive transport of a particle in spatially correlated random energy landscape having exponential density of states has been considered. We exactly calculate the diffusivity in the nondispersive quasi-equilibrium transport regime and found that for slow decaying correlation functions the diffusivity becomes singular at some particular temperature higher than the temperature of the transition to the true non-equilibrium dispersive transport regime. It means that the diffusion becomes anomalous and does not follow the usual $\propto t^{1/2}$ law. In such situation the fully developed non-equilibrium regime emerges in two stages: first, at some temperature there is the transition from the normal to anomalous diffusion, and then at lower temperature the average velocity for the infinite medium goes to zero, thus indicating the development of the true dispersive regime. Validity of the Einstein relation is discussed for the situation where the diffusivity does exist.\end{abstract}

\maketitle

\section{Introduction}

Diffusion of a particle in the random energy landscape serves as a useful model of various physical processes, such as diffusion in amorphous materials, electric conductivity in disordered semiconductors, dynamics of polymer molecules and others. Hopping conductivity of amorphous materials was one of the first examples of the application of this concept. For example, in the simplest 1D case the long time behavior of the carriers' packet $n(x,t)$ for the particular realization of the random energy landscape $U(x)$ may be described by the diffusion equation
\begin{equation}\label{diffusion}
\frac{\partial n}{\partial t}=D_0 \frac{\partial}{\partial x}\left[\frac{\partial n}{\partial x}+\frac{1}{kT}\left(\frac{\partial U}
{\partial x}-eE\right)n\right].
\end{equation}
Here $D_0$ is a bare diffusion coefficient in the absence of disorder and $E$ is the applied electric field. Dynamics of the carrier is a mixture of the diffusion and the drift induced by the action of the applied field. Characteristic features of the dynamics are governed by the density of states (DOS) $g(U)$ and statistical characteristics of the random energy landscape $U(x)$.

In many amorphous semiconductors the typical feature is an exponential tail of the DOS
\begin{equation}\label{exp-DOS}
g(U)\propto e^{U/U_0},\hskip10pt U < 0.
\end{equation}
This feature to a very large extent determines charge transport properties of the materials. In inorganic semiconductors exponential DOS starts at some energy $U_c$ called transport edge and separating localized and extended states, though in some cases this energy could be located in the region of localized states \cite{Schiff:5265}. If states above $U_c$ are delocalized, then it is natural to consider charge transport using the multiple trapping model \cite{Noolandi:4466}, and in the opposite case a hopping transport is the more appropriate model. Recently it was suggested that in some amorphous organic semiconductors there is an exponential tail of the DOS \cite{Vissenberg:12964,Tachiya:085201,Street:165207,Schubert:024203,
Frost:255,Kirchartz:165201,Volker:195336,Foertig:115302,Street:075208,Keivanidis:734}.
Even in perfect organic crystals, where the bands are very narrow, charge transport at the room temperature is dominated by hopping due to thermally induced dynamic disorder and polaronic effects \cite{Troisi:086601,Troisi:2347,Ciuchi:245201,Fratini:2292,Gershenson:973}. In amorphous organic semiconductors significant static disorder leads to the total localization of all states. For such materials the charge transport is exclusively provided by the hopping mechanism.

Hopping charge transport for the exponential DOS has been studied extensively for many years and is considered as rather well studied area \cite{Monroe:146,Rudenko:177,Baranovskii:283}.
The most fundamental fact about such transport is that for low temperature $kT/U_0 < 1$ carriers do not attain a quasi-equilibrium state with constant average velocity but the carrier velocity monotonously decreases with time and, hence, with the thickness $L$ of the transport layer as
\begin{equation}\label{disp-L}
v_L\propto 1/L^{\frac{1}{\alpha}-1}, \hskip10pt \alpha=kT/U_0,
\end{equation}
indicating the transition to the non-equilibrium dispersive transport regime, while for $kT/U_0 > 1$ a quasi-equilibrium regime with constant velocity $v$ eventually develops \cite{Rudenko:177,Arkhipov:189}. There is a general agreement that for the high temperature $kT/U_0 > 1$ the long time behavior is accurately described by the normal diffusive transport.

Nonetheless, in all previous studies a very important ingredient of the random energy energy landscape has been omitted, namely the possible spatial correlation of random energies. Recently we considered the behavior of the average particle velocity $v$ for the spatially correlated exponential DOS and its dependence on temperature and driving force strength \cite{Novikov:24504}. Exact analytic solution has been obtained for the quasi-equilibrium nondispersive regime $kT/U_0 > 1$ and an approximate approach has been suggested for the dispersive regime $kT/U_0 < 1$. Here we are going to extend our consideration to the calculation of the diffusivity $D$, exclusively concentrating our attention on the nondispersive case. It turns out that this case is not so trivial for the spatially correlated energy landscape.

For the sake of simplicity we use the exact exponential DOS with sharp cut-off at $U=0$
 \begin{equation}\label{exp-DOS1}
g(U)=\frac{N_0}{U_0}e^{U/U_0},\hskip10pt U < 0,
\end{equation}
where $N_0$ is the total concentration of hopping sites and $g(U)=0$ for $U > 0$. We mostly consider the temperature being close to $U_0/k$, where particle spends almost all time in the deep states $U\ll -U_0$. For this reason the exact form of the DOS for $U > 0$ is not important. More thorough discussion may be found in Ref. \onlinecite{Novikov:24504}.

For the 1D case Parris et al. derived a general formula for the diffusivity of the charged particle moving in the arbitrary potential $U(x)$ which is periodic with the period $L$ \cite{Parris:5295,Parris:2803}. In the limit $L\rightarrow\infty$ this formula provides the possibility to obtain the relation for the diffusivity statistically averaged over realizations of $U(x)$
\begin{eqnarray}
\label{D}
&D=D_0\frac{\partial v}{\partial v_0}+\Delta D,\hskip10pt v=\frac{D_0}{\int\limits_0^\infty dx e^{-\gamma x}Z_2(x)},\\
\label{rr}
&\Delta D=\frac{v^3}{D_0^2}\int\limits_0^\infty dy dz e^{-\gamma(y+z)}G(y,z),\hskip10pt \gamma=v_0/D_0,\\
\label{lin}
&G(y,z)=\lim\limits_{L\rightarrow\infty}\frac{1}{L}\int\limits_0^L dx x \left[Z_4(x,y,z)-Z_2(y)Z_2(z)\right],\\
\label{z2z4}
&Z_2(x)=\left<e^{\left[U(x)-U(0)\right]/kT}\right>,\hskip10pt
Z_4(x,y,z)=\left<e^{\left[U(x+y)-U(x)+U(0)-U(-z)\right/kT}\right>,
\end{eqnarray}
here the angular brackets mean the average over realization of $U(x)$, $\mu_0$ and $v_0=\mu_0 E$ are the carrier mobility and velocity for the system without disorder. Hence, the task to calculate the diffusivity is essentially shifted to the calculation of the correlation functions $Z_2(x)$ and $Z_4(x,y,z)$. If the second term in \eq{D} is zero, then this equation is, in fact, the modified Einstein relation (mER) \cite{Parris:5295,Parris:2803}, albeit written using the  carrier velocity $v_0$ instead of the applied electric field $E$. The simple Einstein relation
\begin{equation}
D=\frac{kT}{e}\mu
\label{sER}
\end{equation}
is not valid in the materials having field-dependent  carrier mobility $\mu$, apart from the limit case $E=0$. We will consider the applicability of the mER for the exponential spatially correlated DOS.

Correlation functions $Z_2(x)$ and $Z_4(x,y,z)$ diverge for low temperature $kT < U_0$ (dispersive transport regime). This can be immediately seen when all coordinates of $Z_2(x)$ and $Z_4(x,y,z)$ go to infinity, because all exponentials in \eq{z2z4} become independent and, for example,
\begin{equation}
Z_2(x\rightarrow\infty)\rightarrow \left<e^{U(x)/kT}\right>\left<e^{-U(0)/kT}\right>=
\frac{1}{1-\varkappa^2},\hskip10pt \varkappa=U_0/kT.
\label{z2inf}
\end{equation}
Correlation function $Z_4(x,y,z)$ has a similar singularity at $\varkappa\rightarrow 1$. This divergence indicates the invalidity of Eqs. (\ref{D}-\ref{z2z4}) connected with the transition to the non-equilibrium dispersive transport regime at $\varkappa > 1$ where the average velocity goes to 0 for the infinite medium. The difficulty to carry out the calculations for the dispersive regime is directly connected with the necessity to consider transport for a transport layer having the finite thickness. Possibility to consider quasi-equilibrium transport in the infinite medium greatly facilitates the calculations.

It is easy to see that the divergence at $\varkappa\rightarrow 1$ is the only singularity of $Z_2(x)$ and $Z_4(x,y,z)$ for the case of spatially uncorrelated disorder where all exponentials in \eq{z2z4} are independent and $v$ and $D$ are well-defined parameters for all temperatures above the transition temperature $T_0=U_0/k$. For this reason there is a general belief that for the nondispersive regime $\varkappa < 1$ transport could be rather accurately described by the diffusion equation with well-defined average velocity $v$ and diffusivity $D$. In this paper we study the nondispersive regime in details and show that for the correlated random energy landscape the picture is more complicated and diffusivity $D$ demonstrates quite nontrivial behavior.

\section{Gaussian representation for the exponentially distributed random energy $U(x)$}
\label{Gauss}

A suitable method to introduce spatial correlation in the exponentially distributed random energy landscape has been suggested in Ref. \onlinecite{Novikov:24504}. There is a well known representation for the exponentially distributed random variable which employs two auxiliary independent identically distributed random Gaussian variables $X$ and $Y$ having zero mean and unit variance $\sigma^2=1$. If \begin{equation}
U=-\frac{1}{2}U_0\left(X^2+Y^2\right),
\label{U-XY_a}
\end{equation}
then $U$ is distributed according to \eq{exp-DOS1} \cite{Devroye-book}. If $X(x)$ and $Y(x)$ are Gaussian random fields with correlation functions $c_X(x)$ and $c_Y(x)$, then the resulting distribution of $U(x)$ is spatially correlated, too. Using this trick the averaging over $U(x)$ is replaced by the averaging over $X(x)$ and $Y(x)$.

This approach gives us the possibility to generated only limited subset of all exponentially distributed random energies $U(x)$, but we can produce the random field $U(x)$ having an arbitrary nonnegative binary correlation function $c_U(x)$ because
\begin{equation}\label{2a-corrU}
c_U(x)=\left<U(x_1) U(x_2)\right>-\left<U\right>^2=\frac{U_0^2}{2}\left[c_X^2(x)+c_Y^2(x)\right],
\hskip10pt x=x_1-x_2.
\end{equation}

We may consider the use of auxiliary variables $X(x)$ and $Y(x)$ as just a technical trick to introduce spatial correlation in the exponential random energy landscape. This trick facilitates the calculations and there is no need to assume any physical meaning for $X$ and $Y$. Yet there is some physical justification for the Gaussian trick as a way to introduce spatial correlation to the exponential distribution, because May et al. demonstrated a particular mechanism to produce the almost exponential distribution in amorphous organic materials based essentially on some modification of \eq{U-XY_a} \cite{May:136401}.

The bivariate distribution for $X$, needed to calculate $Z_2(x)$, has the form
\begin{equation}\label{2a-PDF_GX}
P_G(X_1,X_2,c_X)=\frac{1}{2\pi\sqrt{1-c^2_X}}\exp\left[
-\frac{X_1^2+X_2^2-2c_X X_1X_2}{2\left(1-c^2_X\right)}\right],
\end{equation}
here $X_i=X(x_i)$, $c_X=c_X(x_1-x_2)$ (the distribution for $Y$ has a similar form) \cite{Feller:book}.
Quadratic form in the exponent of \eq{2a-PDF_GX} is positively defined if $|c_X|<1$. For $x=0$ we have $c_{X,Y}(0)=\sigma^2=1$ and for $x\rightarrow\infty$ $c_{X,Y}(x)\rightarrow 0$.
Correlation function $Z_2(x)$ has been calculated previously \cite{Novikov:24504} and is equal to
\begin{equation}
Z_2(x)=\frac{1}{1-\varkappa^2\left[1-c^2(x)\right]}
\label{Z}
\end{equation}
for $\varkappa < 1$, here and later we consider the simplest case  $c_X(x)=c_Y(x)=c(x)$, possible generalizations may be found in Ref. \onlinecite{Novikov:24504}. We assume also that our system is isotropic, so $c(x)=c(-x)$. To calculate $Z_4(x,y,z)$, we need the analogous 4-point distribution
\begin{equation}\label{P4}
P_4(X_i,c)=\frac{1}{(2\pi)^2\sqrt{\det \textbf{G}^{-1}}}\exp\left[
-\frac{1}{2}\sum\limits_{i,j=1}^4 X_i X_j G_{ij}\right],
\end{equation}
where the elements of the symmetric $4\times 4$ matrix $\textbf{G}^{-1}$ are $G^{-1}_{ij}=c(x_i-x_j)$. Here $x_1=x+y$, $x_2=x$, $x_3=0$, and $x_4=-z$. Performing averaging over $X_i$, $Y_i$ using distribution (\ref{P4}) we obtain for the correlation function $Z_4(x,y,z)$
\begin{equation}
Z_4(x,y,z)=\frac{1}{\det \textbf{T}}, \hskip10pt \textbf{T}= \textbf{G}^{-1}\widetilde{\textbf{G}},
\label{Z4}
\end{equation}
where matrix $\widetilde{\textbf{G}}$ differs from $\textbf{G}$ only in diagonal elements $\widetilde{G}_{ij}=G_{ij}+(-1)^{i-1}\varkappa\delta_{ij}$.

Matrix $\textbf{G}^{-1}$ has the structure
\begin{equation}
\textbf{G}^{-1}=
\begin{pmatrix}
    \textbf{A}       & \textbf{C} \\
    \textbf{C}'       & \textbf{B}
\end{pmatrix},
\label{Ginf}
\end{equation}
where $\textbf{A}$, $\textbf{B}$, and $\textbf{C}$ are $2\times 2$ matrices, $\textbf{C}'$ is the transpose of $\textbf{C}$, and only elements $C_{ij}$ do depend on $x$ and go to 0 for $x\rightarrow \infty$. We may write
\begin{equation}
\textbf{T}=\textbf{G}^{-1}\widetilde{\textbf{G}}=\textbf{I}+\varkappa\textbf{G}^{-1}\textbf{D}=
\textbf{I}+\varkappa
\begin{pmatrix}
    \textbf{A}       & \textbf{C} \\
    \textbf{C}'       & \textbf{B}
\end{pmatrix}
\begin{pmatrix}
    \textbf{d}       & \textbf{0} \\
    \textbf{0}       & \textbf{d}
\end{pmatrix},
\label{Ginf2}
\end{equation}
here $\textbf{d}$ is a $2\times 2$ diagonal matrix having elements $d_{ii}=(-1)^{i-1}$. The final explicit form of matrix $\textbf{T}$ is presented in Appendix \ref{appA}.

\section{Breakdown of the normal diffusive transport in the nondispersive transport regime}

\subsection{General consideration}

Calculation of $\det\textbf{T}$ is carried out in the Appendix \ref{appA}. The result (\ref{detT}) shows that
\begin{equation}
Z_4(x,y,z)-Z_2(y)Z_2(z)\simeq O\left(c^2(x)\right)
\label{y->infty}
\end{equation}
for $x\rightarrow\infty$. Hence, the integral in \eq{lin} grows faster than $L$ and the function $G(y,z)$ diverges if $c(x)\propto 1/x^n$ with $n < 1/2$ for $x\rightarrow\infty$. We see that for slowly decaying correlation function $c(x)$ the diffusivity does not exist, spreading of the photocurrent transients is anomalous and does not follow the law $\propto t^{1/2}$. For the exactly critical correlation function $c(x)\propto 1/x^{1/2}$ the function $G(y,z)$ is finite and diffusivity does exist but the mER is not valid. In this model the breakdown of the mER is a signal of the emerging development of the anomalous diffusive regime.

It turns out that the superlinear growth of the integral (\ref{lin}) is not the only mechanism of the breakdown of the normal diffusive transport (we will call it the mechanism I). Another possibility is the divergence of the correlation function $Z_4(x,y,z)$ for $\varkappa < 1$ (the mechanism II).  Let us consider this possibility. We start with the consideration of the particular case of the exponential correlation function $c(x)=\exp(-x/a)$, where all terms $c(x+y)$, $c(y+z+x)$, and $c(z+x)$ could be expressed through $c(x)$, $c(y)$, and $c(z)$ as $c(x+y)=c(x)c(y)$ etc. Result of the Appendix \ref{appA} gives for that case
\begin{equation}
\det \textbf{T} = \left[1-\varkappa^2+\varkappa^2 c^2(y)\right]
\left[1-\varkappa^2+\varkappa^2 c^2(z)\right]+\varkappa^2 \left(1-\varkappa^2\right)c^2(x)\left[1-c^2(y)\right]\left[1-c^2(z)\right].
\label{exp-c}
\end{equation}
We see that in this case $\det\textbf{T}$ becomes equal to 0 (and, hence, $Z_4(x,y,z)$ becomes singular) only at $\varkappa=1$, exactly at the critical point of the transition to the dispersive non-equilibrium transport regime.

\begin{figure}[tbp]
\includegraphics[width=3in]{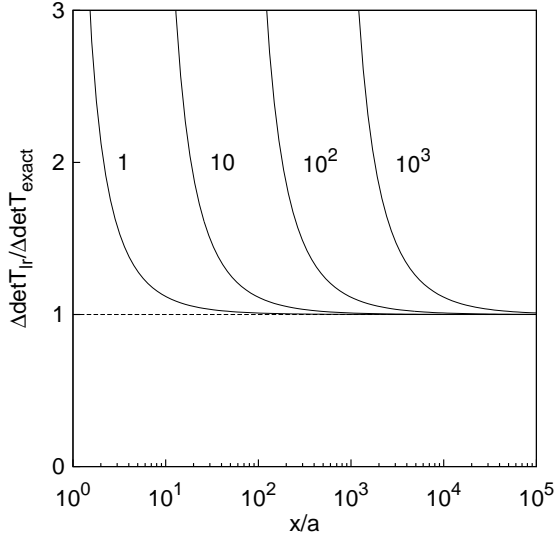}
\caption{Test of the applicability of the simple approximation (\ref{long-c}) for $\det \textbf{T}$. Solid lines show the behavior of $\Delta \det \textbf{T}_{\rm lr}(x,y,y)/\Delta \det \textbf{T}(x,y,y)$ where $\Delta \det \textbf{T}=\det \textbf{T}(x,y,y)-Z_2^{-2}(y)$ for various values of $y/a$ (indicated near the corresponding curve) and for correlation function $c(x)=1/\left[(x/a)^{1/2}+1\right]$ ($\varkappa=0.5$).}  \label{test-lr}
\end{figure}

The situation becomes different for the the case of very long range correlation, where for $x\rightarrow\infty$ we can neglect $y$ and $z$ in $c(x+y)$ etc., assuming $c(x+y)\approx c(x)$ because \eq{rr} shows that the typical values of $y$ and $z$ are $\simeq 1/\gamma$. In this situation $C_{ij}\approx c(x)$ for all $i,j$ and for $x\rightarrow\infty$
\begin{equation}
\det \textbf{T}_{\rm lr} \approx \left[1-\varkappa^2+\varkappa^2 c^2(y)\right]
\left[1-\varkappa^2+\varkappa^2 c^2(z)\right]- 4\varkappa^4 c^2(x)\left[1-c(y)\right]\left[1-c(z)\right].
\label{long-c}
\end{equation}
Validity of the approximation (\ref{long-c}) is demonstrated in Fig. \ref{test-lr}. We immediately see that for the long range disorder $\det \textbf{T}$, contrary to the case of the exponential $c(x)$, depends on the sign of the auxiliary correlation function. Hence, for the critical correlation function $c(x)\propto 1/x^{1/2}$, where $\Delta D$ is nonzero, the diffusivity is sensitive to the sign of $c(x)$, in striking contrast to the average velocity $v$.

\begin{figure}[tbp]
\includegraphics[width=3in]{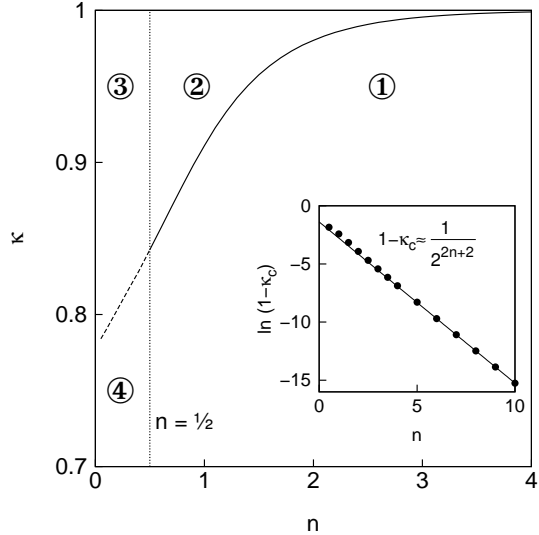}
\caption{Different domains in $n-\varkappa$ plane for the power law correlation function (\ref{phase-power}). Finite diffusivity does exist in the domain 1, while in the domains $2 - 4$ the anomalous diffusion takes place (see more detailed explanation in the text). Insert shows the result of the analytical calculation of the dividing line $\varkappa=\varkappa_c(n)$ between domains 1 and 2 according to \eq{pl-min} (solid line) and numerical calculation of $\varkappa_c(n)$ (points). We neglect here the difference between $\varkappa_c+1$ and 2.}  \label{kappa-power-law}
\end{figure}

In addition, \eq{long-c} hints for the possibility for $\det \textbf{T}$ becomes negative for $\varkappa < 1$, while for the exponential function $c(x)$ this is not possible. Nonpositivity of $\det \textbf{T}$ means that the correlation function $Z_4(x,y,z)$ diverges and, hence, the diffusivity does not exists, too (this is previously mentioned mechanism II). Fig. \ref{kappa-power-law} shows the phase diagram for the power law correlation function
\begin{equation}
c(x)=\frac{1}{(x/a)^n+1},
\label{phase-power}
\end{equation}
i.e. the regions in the plane $n-\varkappa$ where the diffusivity does exist or not exist. In Fig. \ref{kappa-power-law} the vertical line at $n=1/2$ separates the domain $n< 1/2$ where $D$ does not exist for any $\varkappa <1$ (because of the superlinear growth of the integral (\ref{lin}), mechanism I). The solid line in Fig. \ref{kappa-power-law} shows numerically calculated critical values $\varkappa_c(n)$, and for a given $n$  $\det \textbf{T}$ is nonpositive for $\varkappa >\varkappa_c(n)$. We define $\varkappa_c$ as a minimal $\varkappa$ for which the equation
\begin{equation}
\det \textbf{T}=0
\label{det0}
\end{equation}
is valid for some $(x,y,z)$. Above the critical line $D$ does not exist because of the divergence of $Z_4(x,y,z)$  (mechanism II). The domain 1 is the region where long time charge transport demonstrates well-defined average velocity $v$ and diffusivity $D$. Above that line in the domain 2 the diffusion of the carriers is anomalous one, though the average velocity still exists. Formally, the line $\varkappa=\varkappa_c(n)$ may be extended to the region $n<1/2$ (it is shown as a broken line here), thus separating this region into two parts, domain 3 and 4. Exact dynamics of the carriers' packet spreading in domains 2-4 and to what extent they differ from one another are the open problems. The very possibility to subdivide the region to the left of the vertical line $n=1/2$ into two domains is problematic.

\begin{figure}[tbp]
\includegraphics[width=3in]{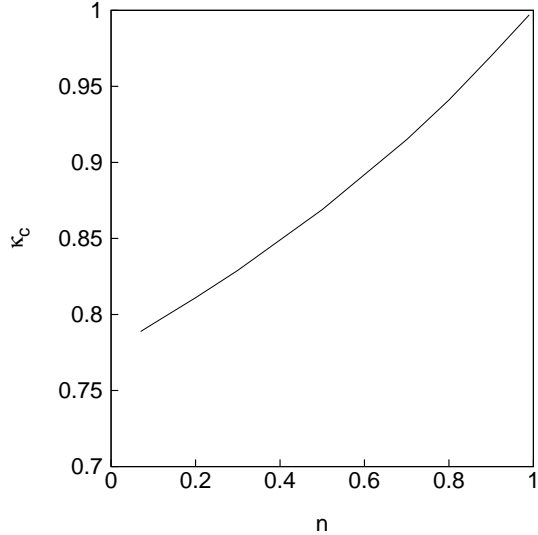}
\caption{Dependence of the numerically calculated critical $\varkappa_c$ on $n$ for the correlation function $c(x)=\exp\left[-(x/a)^n\right]$ (solid line). Above that line the function $Z_4(x,y,z)$ diverges.}  \label{kappa-exp}
\end{figure}

Divergence of $Z_4(x,y,z)$ for $\varkappa < 1$ is not a unique feature of the power law correlation function and also takes place for faster decaying correlation functions $c(x)$, for example, such as $c(x)=\exp\left[-(x/a)^n\right]$ with $n < 1$, see Fig. \ref{kappa-exp}. For the pure exponential correlation function with $n=1$ $Z_4(x,y,z)$ always converges for $\varkappa <1$ (see \eq{exp-c}). Exponential function separates correlation functions permitting the development of the anomalous diffusion for some $\varkappa < 1$ from those giving only the normal diffusion: correlation functions decaying faster then the exponential one, always have the finite $D$ for $\varkappa <1$. This statement can be easily verified analytically for the finite range correlation function $c(x)=\theta(a-x)$ or numerically for the Gaussian correlation function $c(x)=\exp\left[-(x/a)^2\right]$.

Numerical evaluation of $\varkappa_c$ reveals also that for every tested correlation function $c(x)$ the solution of \eq{det0} for the minimal (critical) $\varkappa$ is achieved for $y\rightarrow\infty$. Note that variable $y$ is unique with respect to this feature, it not valid for $z$ and $x$ which remain finite. This particular property is related to the structure of $Z_4(x,y,z)$: in \eq{z2z4} the term $U(x+y)$ enters the exponent with the plus sign. The random energy $U(x+y)$ is negative in all cases and for the divergence of $Z_4(x,y,z)$ its most favorable value is $U(x+y)=0$ irrespective of all other $U$ terms. Absence of correlation between $U(x+y)$ and other $U$ terms facilitates the fulfilment of this requirement and the easiest way to keep $U(x+y)=0$ without affecting other terms is to set $y\rightarrow\infty$.

\begin{figure}[tbp]
\includegraphics[width=3in]{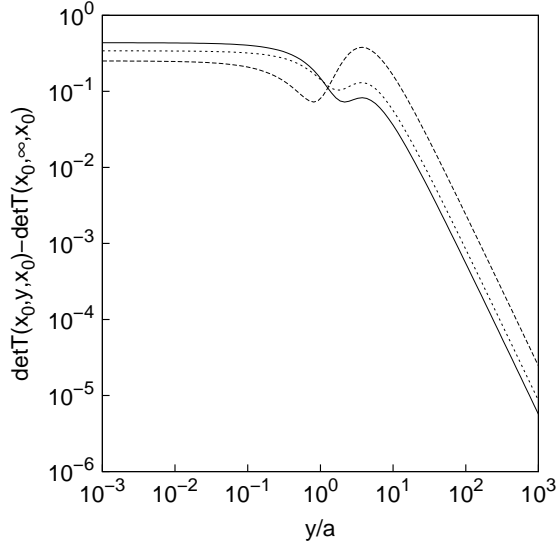}
\caption{Difference between $\det\textbf{T}(x_0,y,x_0)$ and $\det\textbf{T}(x_0,y\rightarrow\infty,x_0)$ where $x_0$ is the optimal distance such as $c(2x_0)=c_{\rm min}$ (here $c_{\rm min}$ is the minimal value of $c(x)$, see explanation for the particular choice of $x_0$ in Appendix \ref{appB}) for the correlation function $c(x)=\left(1-x/a\right)/\left[1+s(x/a)^2\right]$ and  $c_{\rm min}$ equal to -1 (solid line), -1/3 (dotted line), and -0.1 (broken line). For every case $\varkappa$ was chosen as the critical $\varkappa_c$ according to \eq{uni1}. Particular value of $c_{\rm min}$ was set by the proper choice of $s$.}\label{w}
\end{figure}

This important observation means that we may set $y\rightarrow\infty$ and $c(y+...)=0$ in \eq{detT} from the very beginning and consider instead the truncated determinant
\begin{eqnarray}
\det \textbf{T}_{\rm tr}=\left(1-\varkappa^2\right)^2+
\varkappa^2(1-\varkappa^2)\left[c^2(z)+c^2(x)\right]
-\varkappa^2(1+\varkappa)^2 c^2(x+z)+\\
\nonumber
+2\varkappa^3(1+\varkappa)c(x)c(x+z)c(z).
\label{red-detT}
\end{eqnarray}
In future for all analytic calculations we will use $\det \textbf{T}_{\rm tr}$. Fig. \ref{w} shows a typical example of the increase of $\det\textbf{T}$ for the finite $y$.

\begin{figure}[tbp]
\includegraphics[width=3in]{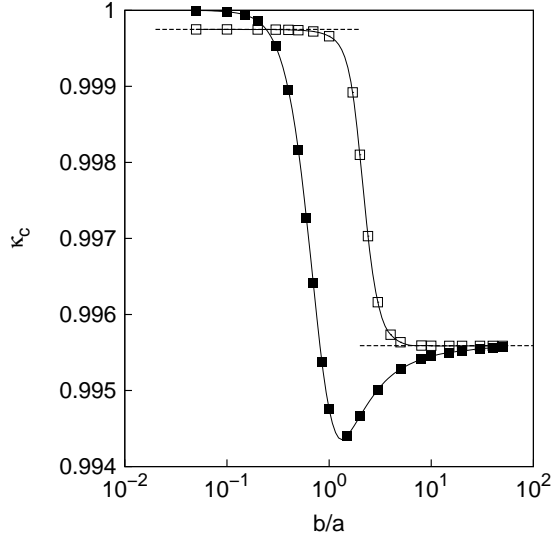}
\caption{Dependence of the critical $\varkappa_c$ on $b/a$ for the correlation functions $c(x)=1/\left[(x/a)^3+\exp\left(x/b\right)\right]$ (filed squares) and $c(x)=1/\left[(x/a)^3+(x/b)^5+1\right]$ (empty squares). Solid lines are provided as a guide for an eye. Broken lines indicates $\varkappa_c$ for $c(x)=1/\left[(x/a)^3+1\right]$ (right side of the plot) and $c(x)=1/\left[(x/b)^5+1\right]$ (left side).}  \label{mix}
\end{figure}

The particular role of the exponential correlation function could be easily understood if we consider the reduced determinant at $\varkappa=1$ (typically, $\varkappa_c$ is rather close to 1). In this case
\begin{equation}
\det \textbf{T}_{\rm tr} = 4c(x)c(z)c(x+z)-4c^2(x+z)=4c(x+z)\left[c(x)c(z)-c(x+z)\right].
\label{particular-exp}
\end{equation}
We assume that $c(x)>0$, hence the sign of $\det \textbf{T}_r$ is determined by the factor $c(x)c(z)-c(x+z)$. For the exponential correlation function $c(x)c(z)-c(x+z)$ is exactly zero. If $c(x)$ decays faster than the exponential one, then the sign of $\det \textbf{T}_{\rm tr}$ is positive, and in the opposite case of the more slow decay it is negative. For $\varkappa=0$ $\det \textbf{T}=1$, so for some $0<\varkappa<1$ there is the zero of the determinant for slowly decaying correlation functions. One should note that the separation of the correlation functions to those decaying more slowly than the exponential ones and decaying faster is considered here not for asymptotics of $c(x)$ for $x\rightarrow\infty$ but for the functional dependence of $c(x)$ in the whole $x$ range. If the function $c(x)$ has the exponential asymptotics but decays differently for moderate $x$, then the critical $\varkappa_c$ still could exist (see Fig. \ref{mix})

\subsection{Power law correlation function $c(x)\propto 1/x^n$}

Use of $\det\textbf{T}_{\rm tr}$ gives us the possibility to carry out an analytical calculation of the critical line $\varkappa_c(n)$ for the power law correlation function (\ref{phase-power}) (more exactly, we assume only that for large $x$ $c(x)\approx (a/x)^n$). Power law correlation functions are typical for amorphous organic semiconductors, though the corresponding models usually provide Gaussian or approximately Gaussian DOS \cite{Novikov:2584,Novikov:2532}. For the approximately exponential DOS considered in Ref. \onlinecite{May:136401} we should expect $c_U(x)\propto 1/x^5$ and $c(x)\propto 1/x^{5/2}$ \cite{Novikov:24504}, though parameter $U_0$ is typically very small and the resulting total DOS in amorphous organic semiconductors is mostly formed by other contributions.

If we assume, in close resemblance to the discussion in Appendix \ref{appB}, that the minimal value of $\det\textbf{T}_{\rm tr}$ is achieving at $x=z$ and for relevant values of $x$ $(x/a)^n\gg 1$ (this is indeed the case for $n\gg 1$, in practice $n\approx 2-3$ is large enough), then $c(x+z)=c(2x)\approx c/2^n$, here $c=c(x)$. Hence,
\begin{equation}
\det \textbf{T}_{\rm tr}\approx \left(1-\varkappa^2\right)^2 +
2\varkappa^2(1-\varkappa^2)c^2-\varkappa^2(1+\varkappa)^2\frac{c^2}{2^{2n}}
+2\varkappa^3(1+\varkappa)\frac{c^3}{2^n}.
\label{det-simple}
\end{equation}
Minimization of this expression with respect to $c$ and subsequent solution of the equation $\det\textbf{T}_{\rm tr}(c_{\rm min})=0$ gives
\begin{equation}
\frac{1-\varkappa_c}{1+\varkappa_c}=\frac{1}{2^{2n+3}}.
\label{pl-min}
\end{equation}
This solution agrees very well with the numerical calculation of $\varkappa_c(n)$ for the full non-truncated $\det\textbf{T}$ (see insert in Fig. \ref{kappa-power-law}).

For the particular case of the power law correlation function $c(x)\propto 1/x^{1/2}$ where for small $\varkappa$ the term $\Delta D$ is finite and nonzero, thus providing the violation of the mER, the diffusivity can be presented in a more appropriate form. It was shown that the approximation (\ref{long-c}) is valid for $x\rightarrow\infty$. Hence, if $c(x)\rightarrow \pm(a/x)^{1/2}$ for large $x$, then
\begin{equation}
G(y,z)= 4\varkappa^4 a Z^2_2(y)Z^2_2(z)\left[1-c(y)\right] \left[1-c(z)\right]
\label{n=1/2}
\end{equation}
and
\begin{equation}
\Delta D = \frac{4\varkappa^4 a v^3}{D_0^2}\left\{\int\limits_0^\infty dy e^{-\gamma y}Z^2_2(y)\left[1-c(y)\right]\right\}^2.
\label{DDn=1/2}
\end{equation}
For the power law $c(x)$ with $n=1/2$ $\varkappa_c=0.8425...$ and is defined by the mechanism II. Nontrivial approximate value of $\Delta D$ could be calculated by the saddle point method for $\varkappa\rightarrow 1$ in close analogy with the calculation of $v$ in Ref. \onlinecite{Novikov:24504} but, unfortunately, this region is unphysical due to the divergence of $Z_4(x,y,z)$. It is interesting to note that \eq{DDn=1/2} does not indicate any singularity at $\varkappa\rightarrow\varkappa_c$ and $\Delta D$ remains finite at that point.

\subsection{Correlation functions $c(x)$ having local extrema}

Such correlation functions demonstrate very remarkable results for the critical $\varkappa_c$ if that value is defined by the mechanism II. Details my be found in Appendix \ref{appB}, here we provide only brief summary.

\begin{figure}[tbp]
\includegraphics[width=3in]{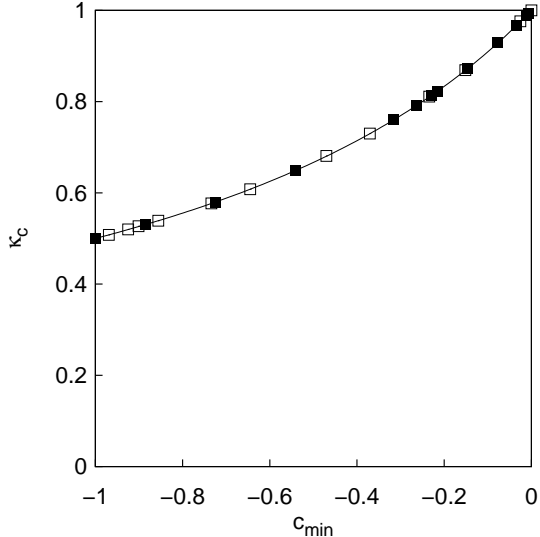}
\caption{Universal dependence of $\varkappa_c$ on the minimal value $c_{\rm min}$ of the correlation function $c(x)$ in the case where $c(x)$ has a minimum with $c_{\rm min} <0 $. Points are results of the numerical calculation of $\varkappa_c$ for  $\det\textbf{T}$ (filled squares for the function $c(x)=\left(1-x/x_0\right)/\left[1+s(x/x_0)^2\right]$, and empty squares  for $c(x)=\exp(-x/x_0)\cos(ax/x_0)$ for various values of $s$ and $a$), while the solid line corresponds to \eq{uni1}. For the second correlation function we considere the deepest (first) minimum of $c(x)$.}\label{uni}
\end{figure}

The simplest case is the correlation function having just one extremum, it has to be a minimum having depth $-1< c_{\rm min}<0$. It turns out that the critical value $\varkappa_c$ is a  universal function of $c_{\rm min}$
\begin{equation}\label{uni1}
   \varkappa_c=\frac{1}{1-c_{\rm min}}
\end{equation}
and does not depend on all other details of $c(x)$, even on the functional form of $c(x)$. In fact, this is true even for $c(x)$ having several extrema if we use the deepest minimum and for all positive maxima $c(x_{\rm max}) <|c_{\rm min}|$ (see Fig. \ref{uni}). If $c(x)$ has an additional positive maximum with height $c_{\rm max} > |c_{\rm min}|$, then the behavior of $\varkappa_c$ is dominated by this maximum but it is a non-universal one.

We should note that the minimal critical value $\varkappa_c$ according to \eq{uni1} is $1/2$. Remember, however, that here we consider the second mechanism of the breakdown of the normal diffusion (i.e., the singularity of $Z_4(x,y,z)$). For the mechanism I (the superlinear growth of the integral (\ref{lin})) the breakdown takes place for the arbitrary small $\varkappa$ if the correlation function decays slow enough.

\section{Experimental evidence for the breakdown of the normal diffusive transport}

\begin{figure}[tbp]
\includegraphics[width=3in]{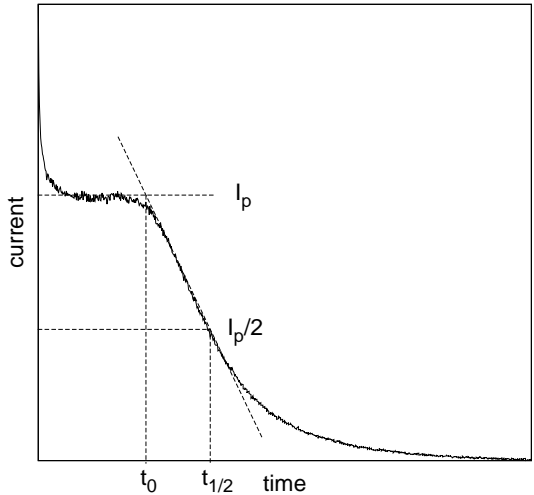}
\caption{Typical photocurrent transient for the nondispersive transport regime (solid line). Broken lines show how to calculate drift times $t_0$ and $t_{1/2}$.}  \label{fig-transient}
\end{figure}

Our main result shows that the nondispersive transport regime for the spatially correlated exponential DOS demonstrates very rich behavior not restricted to the trivial case of the normal diffusion. For various types of the correlation function $c(x)$ the diffusivity does not exist for sufficiently low temperature (though still in the region $U_0/kT < 1$) and spreading of the carriers' packet does not follow usual diffusive law $\Delta x\propto t^{1/2}$. This observation gives a possibility to provide an experimental test of the transition to the anomalous diffusion using the time-of-flight experiment in amorphous organic semiconductors. For the quasi-equilibrium nondispersive transport a typical photocurrent transient demonstrates initial drastic decrease of the current, associated with the spatial and energetic relaxation of the carriers, then development of the plateau of the current where carriers move with the constant average velocity to the opposite electrode, and then a sharp drop of the current when carriers begin to arrive at the collecting electrode (see Fig. \ref{fig-transient}). There is a standard parameter
\begin{equation}\label{W}
    W=\frac{t_{1/2}-t_0}{t_{1/2}},
\end{equation}
which is typically used to describe the spreading of the carrier packet, here $t_{1/2}$ is a time for the transient to decay to the half of the plateau value $I_p$ and $t_0$ is the time of crossing the asymptotes drawn to the plateau and the decaying tail of the transient \cite{Bassler:15,Borsenberger:9,Borsenberger:4289,Borsenberger:967}. For the normal diffusive transport $W\propto \frac{\Delta x}{vt}\propto t^{-1/2}\propto L^{-1/2}$, here $L$ is the thickness of the transport layer. Hence, the investigation of the thickness dependence of $W$ gives a direct possibility to discriminate between the normal diffusive and anomalous nondispersive transport: if $W(L)\propto L^{-1/2}$, then we have a diffusive behavior, and if $W(L)$ decays more slowly, then the diffusivity does not exist and we have the case of the anomalous diffusion.

Amorphous organic materials having localized states where there is an experimental evidence for the existence of the exponential tails of the DOS are best suited for a test of the transition to the anomalous diffusion. Experimental data gives for such materials $T_0=U_0/k$ in the range from 380K to 750K \cite{Vissenberg:12964,Street:165207,Schubert:024203,Frost:255}. This means that to compare our results for the nondispersive transport with the experimental data one have to carry out experiments at rather high temperature. At the moment there are no data for the thickness dependence of $W$ for such materials. Certainly, organic materials having $T_0\simeq 600-700$K cannot be used for the experimental test of our results, the temperature $T_0$ being far too high and we should expect a total thermal decomposition of such organic materials at much more low temperature. Yet materials with $T_0\simeq$400K could be used to study the nondispersive transport in general and dependence $W(L)$ for this transport regime in particular.

Our previous study of the behavior of the average carrier velocity $v$ in spatially correlated exponential random energy landscape shows that nonzero $v$ does exist for any generally acceptable correlation function for $T > T_0$ \cite{Novikov:24504}. Significant variation of $v$ with variation of electric field suitable for a reliable test of the theory could be observed only for temperature pretty close to $T_0$, certainly not far away from $T_0$ than $\simeq 0.01\hskip2pt T_0$. To perform the measurements while keeping the temperature constant with high accuracy is a major obstacle for the experimental test of the theory  (especially taking into account the possible mesoscopic inhomogeneity of the amorphous sample leading to some variation of $T_0$ across the sample). Transition to the anomalous diffusion could occur at the temperature significantly higher than $T_0$, thus providing, in this respect, a less demanding method to test the effects of spatial correlation. At the same time, a very essence of the suggested method requires to work with several experimental samples having different thickness of the transport layer. There is a pressing need to keep the structure of different samples as close as possible, again avoiding mesoscopic inhomogeneities in order to provide a reliable comparison.

At last, we should note that in this paper we consider the behavior of $D$ in the infinite medium. Possible experimental tests are directly related to the finite thickness of transport layers and, hence, the drift time should be much greater than the time of the formation of the final well-developed anomalous diffusion regime. Typical relaxation time for the carrier motion in the correlated exponential DOS is unknown at the moment.

\section{Remarks and conclusions}

We studied the transport of carriers in spatially correlated exponential DOS and found that for a wide range of correlation functions, decaying more slowly than the exponential ones, the diffusivity does not exist for temperatures well above the temperature $T_0$ of the transition to the true non-equilibrium dispersive transport regime. Breakdown of the normal diffusion is caused by the singularity of the term $\Delta D$ in \eq{D}. This term plays a very particular role. Indeed, for almost all correlation functions $c(x)$ $\Delta D$ serves not as a some regular contribution to the total diffusivity but as a trigger, having essentially just two values, zero and singularity. If $\Delta D=0$ then, naturally, it provides no contribution to the total $D$, and if $\Delta D$ is singular, then the normal diffusive transport does not exist and we have the case of the anomalous diffusion. The particular type of the spreading of the carriers' packet and its possible dependence on the form of $c(x)$ will be studied in future.

Transition to the anomalous diffusion could be caused by the action of two independent mechanisms, one is related to the superlinear growth of the integral (\ref{lin}), and the second is related to the singularity of the correlation function $Z_4(x,y,z)$. For the correlation function $c(x)$ decaying more slowly than $\propto 1/x^{1/2}$ the first mechanism provides the development of the anomalous diffusion at arbitrary high temperature. Hence, in semiconductors having very slowly decaying spatial correlations the nondispersive regime is a very peculiar one because the anomalous diffusion occurs at any temperature. Related to the effect of the first mechanism is the invalidity of the modified Einstein relation for the particular case of the auxiliary correlation function $c(x)\propto 1/x^{1/2}$ for $x\rightarrow\infty$, though the normal diffusion still takes place in this case. The second mechanism could lead to the development of the the anomalous diffusion only for $T < 2\hskip2pt T_0$ but for more faster decaying correlation functions. Inequality $T_c < 2\hskip2pt T_0$ is the absolute global boundary for the second mechanism, while the transition temperature $T_c=U_0/\varkappa_c$ for the particular amorphous semiconductor is determined by the actual functional form of the correlation function $c(x)$. If the actual breakdown mechanism is the second one, then the dispersive regime develops in two stages: first, with the decrease of the temperature at $T=T_c$ there is the transition from the normal to anomalous diffusion and then at $T=T_0 < T_c$ the average carrier velocity for the infinite medium goes to zero, thus indicating the development of the full dispersive regime. This new scenario, as well as the very nature of the correlated energy landscape in amorphous organic materials, should have significant implications for the fast developing area of organic electronics, especially if we consider devices operating at different temperatures. Two-stage scenario is reminiscent of the breakdown of the normal diffusion in 1D random force model where the applied driving force $F$ plays the analogue of temperature \cite{Bouchaud:285,Bouchaud:127}. At some critical value $F_{c1}$ there is a transition from the normal diffusion to anomalous one and then, at some $F_{c2}< F_{c1}$ the average velocity for the infinite medium goes to 0.

Study of the nondispersive charge transport, namely the thickness dependence of the parameter $W$ (\ref{W}), in amorphous organic semiconducting materials demonstrating the exponential tail of the DOS and having rather low $T_0\approx 400$K \cite{Vissenberg:12964,Street:165207,Street:075208,Keivanidis:734} should be a promising way to test our predictions. Long range spatial correlation of the random energy landscape is typical for amorphous organic materials.

Future development of the study of the diffusive transport may include the investigation of the phenomenon in the multidimensional case. The method suggested in Section \ref{Gauss} is not restricted to the one-dimensional case and open a possible route to study charge transport for the spatially correlated exponential DOS in many dimensions.    Possible extension to the dispersive transport regime should be studied as well. Particular interest presents the consideration of the diffusive transport in the random exponential energy landscape which cannot be modelled using the Gaussian representation (\ref{U-XY_a}).

Another possible route of development is an extension of the suggested approach to other functional types of DOS. While the main body of the DOS in organic glasses typically has the Gaussian form, far asymptotics of the DOS are expected to decay according different laws \cite{Novikov:164510}. Charge transport is determined by the far tail of the DOS at low temperature, and the the mobility and diffusivity field dependence is governed by the interplay of the functional form of the DOS and nature of spatial correlation. For low $T$ we may expect dependences which differ from those formed by the main body or not so far tails of the DOS and observed for higher temperature.

\section*{Acknowledgements}
Financial support from the Russian Science Foundation grant 15-13-00170 is acknowledged.

\appendix 

\section{Calculation of $\det \textbf{T}$}
\label{appA}

Correlation function $Z_4(x,y,z)$ is the inverse of the determinant of matrix $\textbf{T}$. Direct calculation using \eq{Ginf2} gives
\begin{equation}
\textbf{T}=
\begin{pmatrix}
    1+\varkappa       & -\varkappa c(y) & \varkappa c(x+y) & -\varkappa c(x+y+z) \\
    \varkappa c(y)       & 1-\varkappa & \varkappa c(x) & -\varkappa c(x+z) \\
    \varkappa c(x+y) & -\varkappa c(x) & 1+\varkappa & -\varkappa c(z) \\
    \varkappa c(x+y+z) & -\varkappa c(x+z) & \varkappa c(z) & 1-\varkappa
\end{pmatrix}
\label{T}
\end{equation}
and $\det \textbf{T}$ is
\begin{eqnarray}
\nonumber
\det \textbf{T}=Z_2^{-1}(y)Z_2^{-1}(z)+
\varkappa^2(1-\varkappa^2)\left[c^2(x+y+z)+c^2(x)\right]-\\
\nonumber
-\varkappa^2(1+\varkappa)^2 c^2(x+z)-\varkappa^2(1-\varkappa)^2 c^2(x+y)+\\
\label{detT}
+2\varkappa^3(1+\varkappa)\left[c(x)c(x+z)c(z)+c(y)c(x+y+z)c(x+z)\right]-\\ \nonumber
-2\varkappa^3(1-\varkappa)\left[c(y)c(x)c(x+y)+c(x+y)c(x+y+z)c(z)\right]-\\ \nonumber
-2\varkappa^4 c(y)c(z)\left[c(x)c(x+y+z)+c(x+y)c(x+z)\right]+\\
\nonumber
+\varkappa^4 \left[c(x)c(x+y+z)-c(x+y)c(x+z)\right]^2.
\end{eqnarray}

\section{Calculation of the critical value of $\varkappa$ for the correlation function $c(x)$ having local extrema}
\label{appB}

First, let us consider the simplest case where $c(x)$ has only one extremum at $x> 0$. Taking into account that $c(0)=1$, $-1\le c(x)\le 1$, and  $c(x\rightarrow\infty)\rightarrow 0$, this inevitably means that this extremum is a minimum with $c(x_{\rm min}) < 0$ and $c(x) < 0$ for $x\rightarrow\infty$.

For $\varkappa\ll 1$ $\det \textbf{T}_{\rm tr}(x,z)$ is positive and then, with growing of $\varkappa$, it could become equal to 0 for some critical $\varkappa_c$ at some point $(x,z)$. At this point we have a system of equations to determine $x$, $z$, and $\varkappa_c$ (here for the sake of shorter notation we use the lower index to indicate the argument of the correlation function)
\begin{eqnarray}\label{extremum}
  \det \textbf{T}_{\rm tr}(x,z) &=& 0 \\
  \label{eq2}
  \frac{\partial \det \textbf{T}_{\rm tr}}{\partial x} &\propto&  (1-\varkappa)c'_x c_x
-(1+\varkappa) c'_{x+z}c_{x+z}+\varkappa c_z\left(c'_x c_{x+z}+c_x c'_{x+z}\right) = 0 \\
  \label{eq3}
  \frac{\partial \det \textbf{T}_{\rm tr}}{\partial z} &\propto&  (1-\varkappa)c'_z c_z
-(1+\varkappa) c'_{x+z}c_{x+z}+\varkappa c_x\left(c'_z c_{x+z}+c_z c'_{x+z}\right)=0
\end{eqnarray}
Close inspection of \eq{red-detT} shows that $x+z=x_{\rm min}$ and $c(x+z)=c_{\rm min}<0$ is favorable for the minimization of $\det \textbf{T}_{\rm tr}$. We assume that the condition $x+z=x_{\rm min}$ is indeed valid at the minimum of $\det \textbf{T}_{\rm tr}$ (we will see that this is case). Actually, here we consider a more general case where $x+z=x_e$ and $x_e$ is the position of some extremum of $c(x)$, so $c'(x_e)=0$ and $c(x_e)=c_e$. In this case Eqs. (\ref{eq2},\ref{eq3}) transform to
\begin{eqnarray}\label{extremum2}
  \label{eq2a}
(1-\varkappa)c_x
+\varkappa c_z  c_e &=& 0 \\
  \label{eq3a}
(1-\varkappa)c_z +\varkappa c_x  c_e &=& 0
\end{eqnarray}
or $1-\varkappa=\pm \varkappa c_e$ and $c_x=\mp c_z$.

\begin{figure}[tbp]
\includegraphics[width=3in]{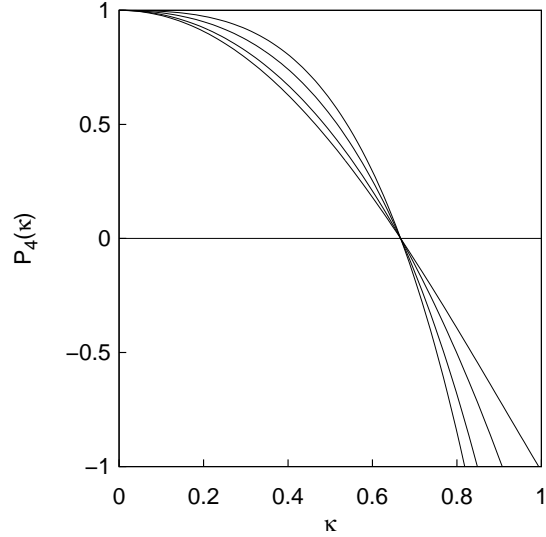}
\caption{This figure demonstrates the independence of the solution of \eq{extremum3} of $c$. Plotted are quartic polynoms $P_4(\varkappa)$ for $c_{\rm min}=-1/2$ and several values of $c$: 1, 0.8, 0.5, and -0.1 (solid lines). All of them cross the abscissa line exactly at the same point thus having the same root $\varkappa_c(-1/2)$=2/3.}  \label{polynom}
\end{figure}

At first, let us consider the case $1-\varkappa=-\varkappa c_e$ and $c_x=c_z=c$. Here the condition $x+z=x_e$ at the minimum of $\det \textbf{T}_{\rm tr}(x,z)$ could be valid  for just one particular value of $\varkappa$
\begin{equation}\label{extremum2a}
\varkappa_c=\frac{1}{1-c_e}.
\end{equation}
We consider the case $\varkappa < 1$, so $c_e$ must be negative. For this case \eq{extremum} becomes
\begin{equation}\label{extremum3}
P_4(\varkappa)=\left(1-\varkappa^2\right)^2 +
2\varkappa^2(1-\varkappa^2)c^2-
\varkappa^2(1+\varkappa)^2 c_e^2
+2\varkappa^3(1+\varkappa)c_e c^2=0.
\end{equation}
Remarkable feature of \eq{extremum3} is the independence of one root of the quartic polynom $P_4(\varkappa)$ of $c$ (see Fig \ref{polynom}). Indeed, $\varkappa_c$ from \eq{extremum2a} fulfills \eq{extremum3} for any $c$ (this can be most easily proved by the direct substitution of $c_e$ expressed in term of $\varkappa_c$ from \eq{extremum2a} to \eq{extremum3}).

Let us consider the matrix of second derivatives of $\det \textbf{T}_{\rm tr}$ in order to prove that the extremum at $c_x=c_z$ is indeed the minimum. For the eigenvalues of the matrix to be positive and extremum being the minimum we must have
\begin{eqnarray}\label{extremum4}
\frac{\partial^2 \det \textbf{T}_{\rm tr}}{\partial x^2}+
\frac{\partial^2 \det \textbf{T}_{\rm tr}}{\partial z^2}=A > 0 \\  \label{eq4b}
\frac{\partial^2 \det \textbf{T}_{\rm tr}}{\partial x^2}
\frac{\partial^2 \det \textbf{T}_{\rm tr}}{\partial z^2}-
\left(\frac{\partial^2 \det \textbf{T}_{\rm tr}}{\partial x \partial z}\right)^2=B > 0.
\end{eqnarray}
Calculation of the derivatives assuming $c'(x+z)=0$ and validity of \eq{extremum2a} gives
\begin{eqnarray}\label{extremum5}
\nonumber
\frac{\partial^2 \det \textbf{T}_{\rm tr}}{\partial x^2}\propto (1-\varkappa)\left(c'_x\right)^2-(1+\varkappa)c_e c''_{x+z}+\varkappa c''_{x+z} c^2, \\
\label{extremum5a}
\frac{\partial^2 \det \textbf{T}_{\rm tr}}{\partial z^2}\propto (1-\varkappa)\left(c'_z\right)^2-(1+\varkappa)c_e c''_{x+z}+\varkappa c''_{x+z} c^2, \\
\nonumber
\frac{\partial^2 \det \textbf{T}_{\rm tr}}{\partial x \partial z}\propto -(1+\varkappa)c_e c''_{x+z}+\varkappa\left(c''_{x+z}c^2+c_e c'_x c'_z\right),
\end{eqnarray}
Here we omit the universal positive coefficient of proportionality $\varkappa^2(1+\varkappa)$. We immediately see that if $c''_{x+z} > 0$, i.e. if we have a minimum of $c(x)$, then all terms in the sum (\ref{extremum4}) are positive and $A > 0$. For parameter $B$ we have
\begin{equation}\label{B}
B\propto -(1-\varkappa)^2 c_{\rm min} c''_{x+z}\left[(c'_x)^2+(c'_z)^2\right]+
\varkappa(1-\varkappa)c''_{x+z} c^2 \left(c'_x+c'_z\right)^2+2\varkappa(1+\varkappa)c''_{x+z} c_{\rm min}^2 c'_x c'_z
\end{equation}
and all terms are positive apart from the last one, its sign being not definite. We can remove this ambiguity setting $x=z=x_{\rm min}/2$, hence automatically $c_x=c_z$ and $c'_x=c'_z$, so the last term becomes positively defined, too.
Finally, we see that for the correlation function $c(x)$ having a negative minimum the choice $x=z=x_{\rm min}/2$ along with $\varkappa$ calculated using \eq{extremum2a} provides a proper solution of Eqs. (\ref{extremum},\ref{eq2},\ref{eq3}), and the extremum of  $\det \textbf{T}_{\rm tr}$ is the minimum. Hence, \eq{extremum2a} with $c_e=c_{\rm min}$ gives a proper value for the critical $\varkappa_c$. The most important feature of \eq{extremum2a} is its universality with value of $\varkappa_c$ being independent of all features of $c(x)$ apart from the depth of its minimum $c_{\rm min}<0$. It is easy to understand that \eq{extremum2a} is valid for $c(x)$ having several minima if we take as $c_{\rm min}$ the depth of the deepest one.

If we consider the corresponding calculation for the second choice $1-\varkappa=\varkappa c_e$ and, hence, $c_x=-c_z=c$ (remember that $x+z=x_e$), we obtain that in this case
\begin{equation}\label{ext6}
\varkappa_c=\frac{1}{1+c_e},
\end{equation}
so $c_e >0$. For the positivity of parameters $A$, $B$ we need $c''_{x+z} < 0$, i.e. we have the maximum of $c(x)$ at $x=x_e$. Equation analogous to \eq{B} now has the form
\begin{equation}\label{B2}
B\propto -(1-\varkappa)^2 c_{\rm max} c''_{x+z}\left[(c'_x)^2+(c'_z)^2\right]-
\varkappa(1-\varkappa)c''_{x+z} c^2 \left(c'_x+c'_z\right)^2+2\varkappa(1+\varkappa)c''_{x+z} c_{\rm max}^2 c'_x c'_z
\end{equation}
and the positivity of the last term is guaranteed if $c'_x c'_z < 0$. Relation $c_x=-c_z$ for some $x$, $z$ such as $x+z=x_{\rm max}$ could, probably, be fulfilled for some functions $c(x)$ but the resulting \eq{ext6} lack the remarkable universality of \eq{extremum2a} where the choice $x=z=x_{\rm min}/2$ automatically provides $c'_x=c'_z$ and positivity of $B$. We should note that the condition $c'_x c'_z < 0$ is the sufficient but certainly not the necessary one for the positivity of $B$ for the case of the correlation function having a maximum. Hence, \eq{ext6} could provide a valid estimation of $\varkappa_c$ for some functions $c(x)$ where $c'_x c'_z > 0$ at the minimum of $\det \textbf{T}_{\rm tr}$, but again this means that the estimation (\ref{ext6}) is nowhere as universal as \eq{extremum2a}.

Comparison of \eq{extremum2a} and \eq{ext6} suggests that for $c(x)$ having the negative minimum and positive maximum with $c_{\rm max} > -c_{\rm min}$ the behavior of $\varkappa_c$ is dominated by the maximum but this behavior is non-universal. Naturally, if $c_{\rm max} < -c_{\rm min}$, then \eq{extremum2a} is valid.

\end{document}